\documentstyle[12pt]{ioplppt}

\newcommand{\bea}{\begin{eqnarray}}
\newcommand{\ba}{\begin{array}}
\newcommand{\eea}{\end{eqnarray}}
\newcommand{\ea}{\end{array}}
\newcommand{\w}{\mbox{\boldmath $\omega$}}
\newcommand{\s}{\sigma}
\newcommand{\bfs}{\mbox{\boldmath $\sigma$}}
\newcommand{\bfxi}{\mbox{\boldmath $\xi$}}
\newcommand{\met}{{\bf g}}
\newcommand{\bfe}{{\bf e}}

\newcommand{\ve}{\varepsilon}
\newcommand{\deq}{\stackrel{\rm def}{=}}
\newcommand{\al}{\alpha}
\newcommand{\be}{\beta}

\makeatletter
\newcommand{\artsectnumbering}{%
\@addtoreset{equation}{section}
\renewcommand{\theequation}{\thesection.\arabic{equation}}}
\makeatother

\begin{document}

\artsectnumbering

\jl{6}
 
\title{Invariant construction of solutions to Einstein's
      field equations -- LRS perfect fluids I\ftnote{7}{Work supported
by the Swedish Natural Science Research Council}}[Construction of 
solutions to Einstein's equations]

\author{Mattias Marklund\ftnote{1}{E-mail address:
mattias.marklund@physics.umu.se}} 

\address{Department of Plasma Physics, Ume{\aa} University, S-901 87
    Ume{\aa}, Sweden}

\begin{abstract}
 The properties of some locally rotationally symmetric
 (LRS) perfect fluid space-times are examined in order to
 demonstrate the usage of the description of geometries in terms of 
 the Riemann tensor and a finite number of its covariant derivatives 
 for finding solutions to Einstein's field equations.  A new method 
 is introduced, which makes it possible to choose the coordinates 
 at any stage of the calculations. Three classes are examined, one with 
 fluid rotation, one with spatial twist in the preferred direction and the
 space-time homogeneous models. It is also shown that there are no LRS
 space-times with dependence on one null coordinate.  Using an extension of 
 the method, we find the full metric in terms of curvature quantities for 
 the first two classes.
\end{abstract} 



\section{Introduction}

Searching for exact and approximate solutions to Einstein's field
equations can be a cumbersome task, and often {\em ad hoc}
mathematical assumptions are used to solve the obtained
equations. This leads to the problem whether or not the solution is
physically interesting, which need not be obvious. It is therefore
convenient if one can
start from physical restrictions, e.g., on a fluid say, and from that
derive a solution to Einstein's field equations.
 
In general relativity the Riemann tensor
and its covariant derivatives is essentially what one can measure,
i.e., those are the physical observables. From the equivalence problem
\cite{Cartan,Karl} we know that given a metric
${\met}=\eta_{ij}{\w}^i\otimes{\w}^j$ in a moving frame, 
the set $R^{p+1} = \{R_{ijkl},...,R_{ijkl;m_1...m_{p+1}} \}$, 
consisting of the Riemann
tensor and a finite number of its covariant derivatives, gives a
complete local description of the manifold ($p+1$ is defined as the
lowest number for which the derivatives of that order are functionally
dependent of those of lower order). This can be used for
classifying metrics, roughly by comparing the different sets for different
metrics \cite{Karl,Brans}.\footnote{The method has been given the name
  {\em Karlhede classification scheme} (and sometimes the Cartan-Karlhede
  classification scheme).}
This naturally leads to the question of
whether or not it is possible to construct the local
geometry (i.e. if we can find some basis and 
  connection 1-forms satisfying Cartan's equations) of a 
manifold starting from some set of elements which we want to be the
Riemann tensor and its covariant derivatives up to some order 
\cite{KarlLind,Brad,KarlBrad}. This task was
pursued in Refs. \cite{KarlBrad,Lic}, and it was shown that one can
always do this if some {\em integrability conditions} (parts of the
commutators, Ricci identities, cyclic and Bianchi identities)
are satisfied by the elements of the set.

The method has not yet been used in an extensive way for finding solutions
to Einstein's field equations.
Therefore, in this series of papers we will study how the method
applies on {\em locally rotationally symmetric} (LRS) perfect fluids. This
is quite a natural choice, since it has a well defined symmetry
property and contains several physically interesting classes, e.g.,
inhomogeneous models and spherically symmetric models of
astrophysical interest (to be studied in Ref. \cite{Marklund}). Also, it
has been studied by other means, thus making a comparison
between different methods possible.

In Sec. \ref{sec:Method} we review the equivalence problem and its inverse
procedure, i.e., construction of solutions of Einstein's equations.

In Sec. \ref{sec:IC} we set up the conditions for a perfect fluid space-time to
have LRS, and we generate the integrability conditions (IC) for these
space-times. Here we introduce a new technique for keeping the choice of
coordinate(s) undetermined through the calculations. 
We classify the space-times into six classes which are
either disjoint, or need some special treatment. In
Secs. \ref{sec:Rotation}-\ref{sec:ST-Homogen} we investigate
the IC for four of the classes.

In Secs. \ref{sec:Isometry}-\ref{sec:Funs} we generate and solve the
isometry algebra over $F(M)$ for two classes, and find the form of the
metric for these.   

We will use the following conventions:
\begin{itemize}
  \item The metric has signature $(+,-,-,-)$.
  \item Uppercase latin indices refers to the frame bundle basis:
    \begin{eqnarray*}
      A,B,...=1,2,...,k, \\
      P,Q,...=k+1,k+2,...,n(n+1)/2,
    \end{eqnarray*}
    and $I,J,...$ are general indices.
  \item Lowercase latin indices denotes components with respect to the
    moving frame on $M$:
    \[
      i,j,...=1,2,...,n.
    \]
    Boldface lowercase latin indices refers to the functionally
    independent objects in $R^{p+1}$:
    \[
      {\bf a},{\bf b},...=1,2,...,k.
    \]
  \item Uppercase greek indices is used for coordinates on the fibers
    of the frame bundle, i.e., the parameters of the orthogonal group:
    \[
      \Upsilon,\Phi,...=1,2,...,n(n-1)/2.
    \]
  \item Lowercase greek indices are used for the coordinates on $M$:
    \[
      \mu,\nu,...=1,2,...n .
    \]
  \item $f_{|i}$ denotes the directional derivative ${\bfe}_i(f)$, where
    $f$ is some function and ${\bfe}_i$ a tangent basis vector.
\end{itemize}

\section{Method}\label{sec:Method}

Here we first give a short review of the equivalence problem and its
solution.
Suppose we are given two $n$-dimensional Riemannian manifolds 
$(M,{\met})$
and $({\tilde M},{\tilde {\met}})$. 
The two metrics are {\em equivalent} on some
subsets ${U}\subset M$, ${\tilde{U}}\subset {\tilde M}$ if
${\met}|_{U} = {\tilde {\met}}|_{\tilde{U}}$. In terms of coordinates, the 
metrics become ${\met} = g_{\mu\nu}\d{x^{\mu}}\otimes\d{x^{\nu}}$ and 
${\tilde {\met}} = 
{\tilde g}_{\mu\nu}\d{\tilde x}^{\mu}\otimes\d{\tilde x}^{\nu}$, and 
two geometries are equivalent if there exists coordinate relations
${\tilde x}^\mu = {\tilde x}^\mu(x^\nu)$, $\mu,\nu = 1,2,...,n$, such that 
\begin{equation}
  {\tilde g}_{\mu\nu} = \frac{\partial x^\sigma}{\partial
    {\tilde x}^\mu}\frac{\partial x^\tau}{\partial 
    {\tilde x}^\nu}g_{\sigma\tau} \ .
\end{equation}
Since it is difficult to find these coordinate
transformations in general, this approach is not effective for
investigating equivalence.

If one uses a {\em moving frame approach}, i.e., one introduces a tangent
basis $\{{\bfe}_i\}$ and its dual $\{{\w}^i\}$, $i = 1,2,...,n$, (and
analogously for ${\tilde M}$), so  
that
\begin{equation} \label{eq:metrics}
            {\met}  =  \eta_{ij}{\w}^i\otimes{\w}^j \ , \qquad
    {\tilde {\met}} =  \eta_{ij}{\tilde{\w}}^i\otimes{\tilde{\w}}^j \ ,
\end{equation}
where $(\eta_{ij})$ is some constant, symmetric and nonsingular
matrix, the equivalence problem can be put in a new guise. From
(\ref{eq:metrics}) we see that ${\tilde {\w}}^i = {\w}^i$ implies 
${\met} = {\tilde {\met}}$. Unfortunately the converse is not
true. If $G$ is the orthogonal group on $M$, we can use some
representation of $G$ to make transformations of the basis while
keeping the tetrad components of the metric fixed\footnote{Actually,
  there still exist the possibility of discrete transformations
  (spatial reflection etc.) but since these are finite in number they
  are rather trivial (at least in our case). We therefore suppress
  this freedom.}, i.e., $\exists$ some $\Lambda\in G$ 
such that $\eta_{kl} = \Lambda^i\!_k\Lambda^j\!_l\eta_{ij}$ (we will
not bother in distinguishing between $G$ and representations of $G$). This
freedom is captured by letting the basis 1-forms be
defined over the {\em frame bundle} $F(M)$, i.e., they depend on both
the coordinates $\{x^{\mu}\}$ over the manifold $M$ and the
parameters $\{\xi^{\Upsilon}\}$, $\Upsilon = 1,2,...,n(n-1)/2$, of
$G$.
Cartan's equations of structure, which defines the {\em connection 1-forms}
${{\w}^i}_j$ and the {\em curvature 2-forms} ${{\bf R}^i}_{j} {\deq}
\frac12{R^i}_{jkl}{\w}^k\wedge{\w}^l$, on a Riemannian manifold reads
  \begin{eqnarray}
    \d{\w}^{i}        =  {\w}^{k} \wedge {{\w}^{i}}_{k} , \label{eq:Cartan1}\\
    \d{{\w}^{i}}_{j}  =  - {{\w}^{i}}_{k} \wedge {{\w}^{k}}_{j} +
    {{\bf R}^{i}}_{j}\ . \label{eq:Cartan2} 
  \end{eqnarray}
with the condition $\eta_{ik}{{\w}^{k}}_{j} + \eta_{jk}{{\w}^{k}}_{i}  =
0$. Since our 1-forms are defined over the frame bundle, the exterior
derivative is defined as $\d {\deq} \d_x + \d_\xi$,
which implies that the connection 1-forms consists of two parts:
\begin{equation}
  {\w}^i\!_j = ^{(1)}\!\!{{\w}^i\!_j} + ^{(2)}\!\!{{\w}^i\!_j} \ ,
\end{equation}
where $^{(1)}\!{{\w}^i\!_j} {\deq} \Gamma^i\!_{jk}{\w}^k$ and
$^{(2)}\!{{\w}^i\!_j} {\deq}
a^i\!_{j\Upsilon}\d\xi^\Upsilon$, for some
$\Gamma^i\!_{jk}$ and $a^i\!_{j\Upsilon}$ determined by Cartan's
equations. Here $\Gamma^i\!_{jk}$ are the Ricci rotation
coefficients. Therefore, when defined over $F(M)$, the connection 
1-forms are linearly independent of the basis 1-forms. They make up
the missing 1-forms for the basis $\{{\w}^I\}$, $I = 1,2,...,n(n+1)/2$, 
of the cotangent space $T_*[F(M)]$, i.e., 
$\{{\w}^I\} {\deq} \{{\w}^i,{\w}^i\!_j\}$. When $\d x^{\mu} =
0$, i.e., along a fibre in $F(M)$, the $^{(2)}\!{{\w}^i\!_j}$ reduce to
the generators ${\mbox{\boldmath $\tau$}}^i\!_j$ of $G$ and they fulfill
$\d_{\xi}{\mbox{\boldmath $\tau$}}^i\!_j = 
-{\mbox{\boldmath $\tau$}}^i\!_k\wedge{\mbox{\boldmath $\tau$}}^k\!_j$.

Using this, one can state the equivalence problem in the following way
\cite{Cartan,Karl}: Two geometries are equivalent 
if and only if $\exists$ relations ${\tilde x}^\mu = {\tilde x}^\mu(x^\nu)$,
${\tilde\xi}^\Upsilon = {\tilde\xi}^\Upsilon(x^\nu,\xi^\Phi)$ such that
${\tilde{\w}}^i({\tilde x}^\mu,{\tilde\xi}^\Upsilon) =
{\w}^i(x^\mu,\xi^\Upsilon)$.

With the collective notation introduced above for the basis
of $T_*[F(M)]$, we can
write Eqs. (\ref{eq:Cartan1}) and (\ref{eq:Cartan2}) as
\begin{equation}
  \d{\w}^I = \frac12 C^I\!_{JK}{\w}^J\wedge{\w}^K \ ,
\end{equation}
where $C^I\!_{JK}$ essentially represents the Riemann tensor. 
In Refs. \cite{Cartan,Karl} it was proved that two metrics ${\met}$ and 
${\tilde {\met}}$ are equivalent if and only if
\begin{eqnarray*}
  C^I\!_{JK}        =   {\tilde C}^I\!_{JK} \ , \\
  C^I\!_{JK|N_1}    =   {\tilde C}^I\!_{JK|N_1} \ , \\
              \qquad   \vdots \\
  C^I\!_{JK|N_1\cdots N_{p+1}}  =  {\tilde C}^I\!_{JK|N_1\cdots
    N_{p+1}} \ ,
\end{eqnarray*}
are compatible as coordinate relations over $F(M)$. 
Here $p+1$ is the lowest derivative order for which 
$C^I\!_{JK|N_1\cdots N_{p+1}}$ is
functionally dependent
on the derivatives of order $<p+1$. This implies that the set 
$$
 C^{p+1} {\deq} \left\{ C^I\!_{JK},C^I\!_{JK|N_1},\cdots,
   C^I\!_{JK|N_1\cdots N_{p+1}} \right\}
$$ 
gives a complete local description of
the manifold. In terms of the Riemann tensor, this set is given by
$R^{p+1} = \{R_{ijkl},R_{ijkl;m_1},\cdots,R_{ijkl;m_1\cdots m_{p+1}}\}$ 
\cite{Karl}.

Since we can achieve a local description of $M$ in terms of $R^{p+1}$,
it can be used for constructing solutions to Einstein's equations in
an invariant way, because we know that two solutions of Einstein's
equations are equivalent if a comparison between their $R^{p+1}$ gives
consistent coordinate relations. Obviously, any set of $R^{p+1}$
elements can not generate 1-forms 
such that Cartan's equations are fulfilled. Some conditions have to be
imposed on the set.
Let $\{I^{\bf a}\}$, ${\bf a} = 1,2, ... ,k \leq n(n+1)/2$, be a
maximal set of functionally independent
objects chosen from $R^{p}$ [this means that the dimension of our
isometry group is $n(n+1)/2-k$]. As our basis we have chosen
$\{{\w}^I\} = \{{\w}^i,{\w}^i\!_j\}$, where $\{{\w}^i\}$ is some moving frame.
In Refs. \cite{KarlBrad,Lic} it was shown that the integrability
conditions for a set $R^{p+1}$ are 
\begin{eqnarray}
  \d^2 I^{\bf a} = \d\left(I^{\bf a}\!_{|J}{\w}^J\right) = 
  0 \ ,\label{eq:int1} \\ 
  \d^2 {\w}^P = \d\left(\frac12C^P\!_{JK}{\w}^J\wedge{\w}^K\right) 
  = 0 \ ,\label{eq:int2} 
\end{eqnarray}
where $P = k + 1, k + 2, ... ,n(n+1)/2$. When these are fulfilled,
we can solve for $\{{\w}^A\}$, $A = 1,2,\cdots,k$, through Cartan's
equations. In Ref. \cite{Lic} a procedure for finding the rest of the
1-forms, based on Ref. \cite{KarlMac}, was described.
If the manifold is without symmetries, i.e., $k = n(n+1)/2$, we see
that we do not need any of Eqs. (\ref{eq:int2}). 
If we impose some symmetry, we reduce the number of functionally 
independent elements in $R^{p+1}$, and Eqs. (\ref{eq:int2})  
have to be added.
 
In practice it is often easier to work in a fixed frame than to let
the components in $R^{p+1}$ depend explicitly on the parameters
${\xi}^{\Upsilon}$ of the orthogonal group [i.e., we choose a cross-section 
of $F(M)$].
Suppose that $R^{p+1}$ only depend on
$x^{\alpha} \, , \, \, \alpha = 1,2,..., l$, and rotations in the
$ab$-planes, $\{ {{}^{a}}_{b} \} = 1,...,m$,
where $l=n-
 \rm{dim(orbits)}$ and $m=n(n-1)/2-\rm{dim(isotropy  \,\, group)}$.
Equation (\ref{eq:int1}) then correspond to \cite{KarlBrad,Lic}
  \begin{eqnarray}
    \fl \d^2 x^{\alpha}     
    = \d(x^{\alpha}\!_{|i}{\w}^i) = 0  & \Leftrightarrow 
    [{\bfe}_k,{\bfe}_l](x^{\alpha}) = 
         -(\Gamma^j\!_{kl} 
         - \Gamma^j\!_{lk}){\bfe}_j(x^{\alpha}) \ ,\label{eq:Comm}\\
    \fl \d {\mbox{\boldmath $\tau$}}^a\!_b = \d \left(
      {{{\w}}^{a}}_{b} - {{\Gamma}^{a}}_{bi} {{\w}}^{i} \right) 
      & \Leftrightarrow R^a\!_{bij} = 2\left[ {\bfe}_{[i}(\Gamma^a\!_{|b|j]}) 
    + \Gamma^a\!_{m[i}\Gamma^m\!_{|b|j]} +
    \Gamma^a\!_{bk}\Gamma^k\!_{[ij]} \right] \ , \label{eq:RieEq} 
  \end{eqnarray}
where $|$ means exclusion from the antisymmetrisation, 
and the rest of the equations that have
to be satisfied are (\ref{eq:int2}) which can be written
  \begin{eqnarray}
    {\bf R}^t\!_j\wedge{\w}^j = 0  & \Leftrightarrow   R^t\!_{[ijk]} = 0 \ ,
    \label{eq:Cycl}\\
    \d{\bf R}^p\!_q + {\bf R}^p\!_k\wedge{\w}^k\!_q - 
    {\w}^p\!_k\wedge{\bf R}^k\!_q = 0 & \Leftrightarrow   
    R^p\!_{q[ij;k]} = 0\ , \label{eq:Bian}
  \end{eqnarray} 
where $t=l+1, l+2,...,n$ and $\{ {{}^{p}}_{q} \}=m+1,m+2,...,n(n-1)/2$.
For general relativity, the usual 1+3 orthonormal frame approach 
(i.e. splitting of space-time with respect to a timelike congruence) often 
makes use of the field equations together with the Jacobi identity for the
tangent basis. Here they are replaced by Eqs. (\ref{eq:RieEq}) and 
(\ref{eq:Cycl}) (we assume that all other symmetry properties of the Riemann 
tensor are fulfilled).  

In Ref. \cite{KarlBrad} it was also shown that instead of using the
full set $R^{p+1}$, one can use the reduced set $S {\deq}
\{ R^i\!_{jkl},\Gamma^a\!_{bk},x^{\alpha}\!_{|k},\eta_{ij}\}$ to describe
the geometry, since $R^{p+1}$ can be constructed from $S$. Note the difference
in that the set $R^{p+1}$ is covariantly defined, while the set $S$
is not (at least in the generic case).

From the results listed above we see that if the manifold
have translational symmetries only, 
we do not need any of the Bianchi identities. When we add some kind of
isotropy we need some of the Bianchi
identities but we can reduce the number of Ricci identities
needed. 

Thus, to construct a geometry given a set $R^{p+1}$ or $S$, we adopt
the following scheme:
\begin{enumerate}
 \item[1.] Determine the set $R^{p+1}$ or $S$, with the help of the IC.
 \item[2.] Determine the ${{\w}}^{A}$. Since we can write $\d I^{\bf a} =
   I^{\bf a}\!_{|I}{\w}^I$, we can invert this relation for $k$ of the
   1-forms, i.e., for $\{{\w}^A\}$. Since the inverse of $I^{\bf
     a}\!_{|A}$ are part of $R^{p+1}$, these functions are determined
   from the IC.
 \item[3.] Derive the isometry group from the projected Cartan's
   equations.
 \item[4.] Finally, determine ${{\w}}^{P}$ as follows:
   \begin{enumerate}
   \item Using the isometry group, solve for the 1-forms in terms
     of some coordinate basis. Note that we can always pose this as a
     boundary value problem for a set of coupled ordinary differential
     equations (see the constructive proof of Lie's Third Theorem in
     Flanders \cite{Flanders}). In most cases, there already exists
     canonical choices tabulated, e.g., in the 3-dimensional case the
     Bianchi classification \cite{RyanShep} and in the 4-dimensional case the 
     work of MacCallum \cite{Mac}. Now, use one of the
     procedures above to make a metric ansatz ${\met}$ by extending
     the 1-forms found to the entire manifold.
    \item From this, calculate the set $R^{p+1}$ and
      compare with the original set, giving consistency 
      equations for the coefficients in the metric ansatz.
   \end{enumerate}
\end{enumerate}

\section{Integrability conditions}\label{sec:IC}
 
LRS perfect fluids have been extensively studied, because they are 
simple in their symmetry and contains a lot of physically interesting examples
(see
Refs.~\cite{Ellis,StewEllis,EllisMac,Mac2,Collins1,BogoNovi,Bogo,Mac3,
Kramer,vanElst}).
 
The energy-momentum tensor is given by
\[
  {\bf T} = ({\mu} + p){\bf u}\otimes{\bf u} - p{\met} \ ,
\]
where ${\bf u}$ is the 4-velocity of the fluid, ${\mu}$ the energy density,
$p$ the pressure and $\met$ the metric.
We choose a comoving Lorentz-tetrad, i.e. we select the tangent basis 
$\{{\bfe}_i\}$ and its dual $\{ {\w} ^{i} \}$, $i=0,...,3$, such that
\begin{equation}
  {\met} = \eta_{ij}{\w}^i\otimes{\w}^j = ({\w}^0)^2 - ({\w}^1)^2 -
  ({\w}^2)^2 - ({\w}^3)^2 \ , \ \ {\w}^0 = {\bf u} \ .
\end{equation}

We have the following definition of LRS~\cite{Lic,StewEllis,vanElst}: \\
\ \\
{\bf Definition}
  {\em A space-time is said to be {\em LRS in a neighborhood
  $U(p)$ of a point $p \in M$} if at every point $q \in U(p)$ there is a
  subgroup of the proper Lorentz group that leaves invariant the Riemann
  tensor and its covariant derivatives, i.e., there
  exists a continuous isotropy group at each point in the
  neighborhood $U(p)$.}
\ \\
  
We rotate the axes so that our symmetry lies in the 23-plane, which
makes our tetrad fixed up to rotations in that plane.\\
\ \\
{\bf Assumptions:} The essential coordinates on the space-time
manifold $M$ are $x^{0}$ and
$x^{1}$, where $x^{0}$ is time-like and $x^{1}$ is
space-like.\footnote{These are elements of $R^p$, or combinations 
  thereof.} We assume that their exterior derivative can be
written\footnote{Since the $x^\alpha\!_{|i}$ are curvature
  quantities from the set $R^{p+1}$, components in the ${\bfe}_2$ 
  and ${\bfe}_3$ directions would break the LRS.} 
 \begin{eqnarray}
  \d x^{0} & = & X{{\w}}^{0} + Y{{\w}}^{1} , \label{eq:transf1} \\
  \d x^{1} & = & x{{\w}}^{0} + y{{\w}}^{1} , \label{eq:transf2}
 \end{eqnarray}
where $X = {x^{0}}_{|0}$, $Y = {x^{0}}_{|1}$, $x =
{x^{1}}_{|0}$ and $y = {x^{1}}_{|1}$. If the coordinates are independent
we must have $Xy-xY\neq0$. Since $x^{0}$ is time-like,
i.e., $X^{2} - Y^{2} > 0$, we must have $X \neq 0$. An analogous argument
gives that $y \neq 0$. From this we also see that if $X = 0$ or $y =
0$, then  $Y = 0$ or $x = 0$ respectively. \\ 
\ \\
In previous papers \cite{KarlLind,Brad,KarlBrad}, the coordinate
have been specified in advance (for
example, the density). A drawback with this procedure has of course
been that one has then 
restricted once attention to models with some special behavior (in the 
previous example of coordinate, we must not
deal with an incompressible fluid). This drawback is eliminated by using
the general structure introduced above. Also, since there often exists
several choices of coordinates in a given situation, this will enable
one to make the `canonical' choice, i.e., the choice that gives the
simplest equations.

The reduced set $S$ contains the Riemann tensor
 \begin{eqnarray*}
   \fl R_{0101} = E-({\mu} + 3p)/6 \ ,& R_{0202} = R_{0303} =
       -E/2 - ({\mu} + 3p)/6 \ , \\
   \fl R_{1212} = R_{1313} =  E/2 - \mu/3 \ , & R_{2323}
       =  -E - \mu/3 \ , \\
   \fl R_{0123} = 2R_{0213} = -2R_{0312} =  H \  ,     
 \end{eqnarray*}
where $E = E_{11}$ and $H = H_{11}$ are components of 
  \begin{eqnarray}
    E_{ij} &{\deq}& C_{ikjl}u^{k}u^{l} \ , \\
    H_{ij} &{\deq}& {\epsilon}_{iklm}{C^{lm}}_{jn}u^{k}u^{n}/2
    \ ,
  \end{eqnarray}
where $E_{ij}$ and $H_{ij}$ are the electric and magnetic parts of the
Weyl tensor respectively (here
${\epsilon}_{iklm}$ is the totally antisymmetric tensor defined by
${\epsilon}_{0123} =1$).
 
To be able to construct a geometry we also need the rotation-coefficients.
Some of these are expressible in the kinematic quantities acceleration $a_{i}$,
expansion $\Theta$, vorticity $\omega_{ij}$ and shear $\sigma_{ij}$ through
 \begin{equation}
  -\Gamma_{0ij} =
  \nabla_j u_i =\omega_{ij}+\sigma_{ij}-h_{ij}\Theta/3+a_{i}u_{j} .
 \end{equation}
The kinematic quantities are defined as \cite{Ehlers}
 \begin{eqnarray*}
 \omega_{ij}  {\deq}  {h_{i}}^{k}{h_{j}}^{l}\nabla_{[l}u_{k]} , \qquad
 & \sigma_{ij}  {\deq} 
 {h_{i}}^{k}{h_{j}}^{l}\left(\nabla_{(l}u_{k)}+h_{kl}\Theta/3 \right), \\ 
 \Theta {\deq}  \nabla_{i}{u^{i}}  , & a_i    {\deq}  u^j\nabla_j u_i  ,
\end{eqnarray*}
and $h_{ij} {\deq} u_{i}u_{j} - {\eta}_{ij}$ is the projection tensor.
The choice of a comoving frame results in all kinematic quantities with a
$0$ index being zero, $a_{0}=\omega_{i0}=\sigma_{i0}$, because of
their orthogonality to the 4-velocity.
Because of the LRS there is only one independent component of
the vorticity, $\omega_{23}=-\omega_{32} {\deq} \omega$, and one
component of the acceleration, $a_{1} {\deq} a$. The shear will,
because it is traceless, look like $(\sigma_{ij}) =
\rm{diag}(0,-2\sigma,\sigma,\sigma)$.The rotation coefficients are then
 \begin{eqnarray*}
      \Gamma_{010} = -a\ , 
    & \Gamma_{011} =  2\sigma + \Theta/3 \ \deq {\al} , \\ 
      \Gamma_{022} = \Gamma_{033} =  -\sigma +
      \Theta/3 \ \deq {\be}, \qquad  
    & \Gamma_{023} = -\Gamma_{032} =  -\omega\ , \\
      \Gamma_{122} = \Gamma_{133} {\deq}  -\kappa\ ,  
    & \Gamma_{123} = -\Gamma_{132} {\deq}  -\lambda \ ,  
 \end{eqnarray*}
where the restriction on the $\Gamma_{12i}$ and $\Gamma_{13i}$ follows
from the LRS requirement.
We do not need $\Gamma_{23i}$ since we have rotational symmetry in the
23-plane. The rotation coefficient $\kappa$ corresponds to the spatial
divergence of the vector field ${\bfe}_1$, while $\lambda$ is the spatial
rotation of the same vector field relative to a triad which is 
Fermi-transported along ${\bf u}$. Thus all rotation coefficients are
covariantly defined because of the LRS.
 
We can observe that we have a 3-dimensional spatial isotropy group
$H_3$ if and only if $E=H=\s=\omega=a=y=x=Y=0$. This also means
that we do not need the $\Gamma_{12i}$ and $\Gamma_{13i}$, and some of
the Ricci identities are redundant. On the
other hand, we need further components of the Bianchi identities
(these will turn out to give no more information). Spherically
symmetric models must have LRS and an $H_3$ acting only at
one point. Therefore, at that point (the `center' of the model) all
quantities listed above must be zero. They then evolve smoothly from
the center, and the evolution is given by the IC.
 
With all this we are ready to let the IC take
explicit form, and we give them in the Appendix. It is interesting to
note that to determine the magnetic part of the Weyl tensor, $H$, we do
not need any evolution or divergence equations. It is algebraically
related to the other quantities in $R^{p+1}$ [see
Eq. (\ref{eq:H})]. This is because we do not need the components of the
Bianchi identities where $H$ appear differentiated (according to the
theorem given in Refs. \cite{KarlBrad,Lic}). In general, space-times
without any isotropy have their Weyl tensors algebraically determined
in terms of the rotation coefficients, their derivatives and the 
coordinate gradients through the Ricci identities. 
This follows as a corollary of
the theorem presented in Refs. \cite{KarlBrad,Lic}.
 
Although they are not needed (since they are contained in the IC),
things will be greatly simplified if one uses the twice contracted
Bianchi-identities:
 \begin{equation}
   {\mu}_{|0} = -({\mu}+p)\Theta, \qquad p_{|1} = ({\mu}+p)a.
\label{eq:contrBian}
 \end{equation}
These can be obtained by applying ${\bfe}_0$ and ${\bfe}_1$ on parts of the 
Ricci identities, and using the rest of the IC.
 
We can compare the curvature description of geometries used here
with the 1+3 threading formalism presented in the paper 
by van Elst and Ellis~\cite{vanElst}, since
their article resembles the first part of this one (for a general discussion of
the 1+3 orthonormal frame approach, see Ref. \cite{Uggla}). Using their
notation, they choose a vector 
$e^{i}$ as their preferred (normalized) space-like vector field, and 
defines the rotation of it in the usual manner. Since the rotation has
to be proportional to $e^{i}$, they have
 $
  {\epsilon}^{ijkl}\left( {\nabla}_{j}e_{k} \right)u_{l} = -ke^{i} \ ,
 $
for some function $k$. 
If we translate this to our notation, i.e., choosing a comoving Lorentz
tetrad, and using $e^{i} = \delta^i\!_1$ 
(using their signature convention) we get
that $k = 2{\Gamma}_{132} = 2{\lambda}$. Further, the
spatial divergence of $e^i$, 
 $
  a_{{\rm vEE}} {\deq} {h^{i}}_{j}{\nabla}_{i}e^{j}
 $
(index by the author), will satisfy $a_{{\rm vEE}} = 2\kappa$.
The procedure of obtaining the relevant equations differs somewhat between 
the 1+3 formalism and the curvature description. In the former, one generate
the Ricci and Bianchi identities together
with the Jacobi identities. Independence of the equations is then 
checked. One obtains a split in the above equations into the evolution 
equations and the constraint equations 
(which contains only spatial derivatives). 
From this, one checks the consistency by taking the covariant time derivative 
of the constraint equations and demand that this derivative vanish.   
In the Bianchi identities, there are equations 
containing the derivatives of $H$. Now, we can from the curvature description
conclude that these equations are redundant, since (because of the LRS) 
these parts of the Bianchi identities are never needed. Although this is is no
big advantage in this case (because in the three cases below the Weyl tensor 
is either algebraically determined or has zero magnetic part) it points at
the fact that the reduction of the number of equations is automatic.   
 
A classification of the different LRS models can be done
using the IC. One essentially uses
Eqs.~(\ref{eq:class1}) and~(\ref{eq:class2}), together with the
causal properties of the coordinates (i.e., that $x^{0}$ is time-like
and $x^{1}$ is space-like).
 
First of all, if the rotation is
nonzero we can use Eqs. (\ref{eq:class1}) and (\ref{eq:class2}) to write
\begin{equation} \label{eq:Nonzerorotrel}
    X=Y\lambda/\omega \qquad {\rm and} \qquad x=y\lambda/\omega
\end{equation}
respectively. Inserting
this into the causal restrictions on the coordinate gradients, we see
that the equations splits into two cases (i) $x = y = 0$ or (ii) $X =
Y = 0$ (there is also the null coordinate case $x^2-y^2=X^2-Y^2=0$, which we
treat later). Case (i),
using Eq.~(\ref{eq:Nonzerorotrel}) to combine Eqs.~(\ref{eq:B})
and~(\ref{eq:A}), gives ${\be}=-\kappa\lambda/\omega$. Inserting this
into Eqs.~(\ref{eq:C})
and~(\ref{eq:D}) gives $({\mu}+p)\kappa\lambda=0$. Discarding the case
${\mu}+p=0$ for the moment, we check $\kappa =0$, and see
that ${\lambda} = 0$, so ${\lambda} = 0$ always holds [if $\kappa =
0$, Eq.~(\ref{eq:E}) gives ${\lambda}{\omega} =
0$]. But this gives $X=0$, which in turn implies $Y=0$, i.e. no dependence
on any coordinates at all. These space-time homogeneous cases is
treated in Section \ref{sec:ST-Homogen}. Performing an analogous manipulation
for (ii) gives $x=0$. But this does not imply $y=0$. Thus we here have a
nontrivial case. From Eq. (\ref{eq:Nonzerorotrel}) we get $\lambda =
0$. Inserting this into Eqs. (\ref{eq:B}) and (\ref{eq:lambdaB})
results in ${\be} = {\al} =0$, i.e., the fluid is shear- and
expansion-free. 
 
Second, we can have $\lambda \neq 0$. From (\ref{eq:class1}) and
(\ref{eq:class2}) we have
\begin{equation} \label{eq:Zerorotrel}
  Y=X\omega/\lambda \qquad {\rm and} \qquad y=x\omega/\lambda \ ,
\end{equation}
respectively. If we insert this into the causal restrictions on the
coordinate gradients, we obtain (as in the first
class) two separate cases: (i) $x=y=0$ or (ii) $X=Y=0$. For (i), using
the relation between $X$ and $Y$, we can combine Eqs. (\ref{eq:lambdaA})
and (\ref{eq:lambdaB}) and get $\kappa=-{\be}\omega/\lambda$. Making
use of this, Eqs. (\ref{eq:C}) and (\ref{eq:D})
gives the result $({\mu}+p){\be}\omega =0$. As before, we assume
${\mu}+p\neq 0$ for the moment. We see that this implies that
$\kappa=0$, and multiplying Eq.~(\ref{eq:E}) with $\omega$ gives $\omega=0$,
i.e. $Y=0$. From Eq.~(\ref{eq:B}) we obtain $a = 0$, which also can be
seen from Eqs.~(\ref{eq:contrBian}). We can proceed in the same manner for
(ii), and obtain $y=0$. But this implies $x=0$ (see {\bf
  Assumptions}), i.e., space-time homogeneous models (see Section
\ref{sec:ST-Homogen}). 
 
Third, if $\omega =0=\lambda$ we can have dependence on both a
time-like and a space-like coordinate in general. This case have zero
magnetic part of the Weyl tensor, which can be seen from Eq.~(\ref{eq:H}).
 
When $\mu+p=0$, the IC implies that $\mu$ and $p$ are constants. Therefore
this is equivalent to vacuum with a cosmological constant, and we will 
from now on assume that $\mu+p\neq0$.

Thus, the IC gives (when ${\mu}+p\neq 0$):
\begin{enumerate}
 \item[1.] ${{\omega} \neq 0}$ (LRS class I) $\Rightarrow$
   $X=Y=x={\lambda}={\sigma}={\Theta}=0$, no time-like dependence.
 \item[2.] ${{\lambda} \neq 0}$ (LRS class III)
    $\Rightarrow$ $x=y=Y=\kappa=a=\omega=0$, no space-like dependence.
 \item[3.] ${{\omega} = 0 = {\lambda}}$ (LRS class II)
    $\Rightarrow$ $H = 0$,
   generally both space-like and time-like dependence.
\end{enumerate}
The classification in the parenthesis is the one given in
Refs.~\cite{Ellis,StewEllis,vanElst}. Three cases which lies somewhat
outside the above classification are 
\begin{enumerate}
  \item[4.] Space-time homogeneous cases.
  \item[5.] Dependence on one null coordinate.
  \item[6.] ${\mu}+p=0$.
\end{enumerate}
The fourth class contains e.g. the G{\"o}del universe and  
the fifth class will be shown to be homogeneous.
 
In this paper we study the classes {1}, {2}, {4} and {5}. 
The third class contains e.g. the
spherical symmetric case, which means that there is an abundance of
models in it. Therefore we treat it separately in a forthcoming paper
\cite{Marklund}.

\section{Nonzero vorticity: $\omega \neq 0$}\label{sec:Rotation}
 
In this case, we see that many of the kinematic quantities are
automatically zero. This of course means that our system of equations
is heavily reduced. It in fact becomes two parts, one consisting of
differential equations, and one consisting of algebraic relations. 
Inserting the algebraic relations into
the differential ones, we get the differential equations
  \begin{eqnarray}
    y{\omega}' & = & (2\kappa - a)\omega , \label{eq:rot1} \\
    y{\kappa}' & = & ({\mu} + p)/2 - a\kappa - \omega^{2} +
                     \kappa^{2} , \label{eq:rot2} \\
    ya'& = & -({\mu} + 3p)/2 + a^{2} + 2a\kappa +
                     2\omega^{2} , \label{eq:Dustrot} \\
    yp'        & = & a({\mu} + p)\ , \label{eq:ContrBian}
  \end{eqnarray}
where $'$ denotes differentiation with respect to $x^1$, 
and the algebraic relations (which can be seen as defining relations)
\begin{eqnarray}
  E & = & -({\mu} + 3p)/3 + 2a\kappa + 2\omega^{2}, \label{eq:ElRot}\\
  H & = & 2(a - \kappa)\omega. \label{eq:MagRot}
\end{eqnarray}

Equations (\ref{eq:rot1})-(\ref{eq:ContrBian}) contain a set of six functions: 
$\{\omega, \kappa, a, y, p, {\mu}\}$. One of them (or a combination, if
preferable) can be chosen as coordinate, which makes $y$ 
algebraically determined in terms of the other functions. If we assume
an equation of state the number of functions reduces to 3 (counting
only those contained in the differential equations), i.e., the same
as the number of equations. Thus, in the generic case, we need only
solve three differential equations, and the rest of the functions are
determined algebraically.

If there are some algebraic constraints (e.g. $E=0$) we can insert
this in the IC. If no equations turns out to be linearly dependent,
and we assume an equation of state, we can use this to generate a
second order differential equation for the equation of state, as will
be seen below. All other quantities are then algebraically determined
in terms of the pressure and density (and possible constants), when 
choosing the pressure as $x^1$.
 
We can observe that for conformally flat
space-times, the constraints on the kinematic quantities gives (when
inserted into the IC) as equation of state ${\mu}+p=0$.

\subsection{Rigid rotation}

If we have rigid rotation, i.e. $\omega' = 0$, Eq. (\ref{eq:rot1})
gives $a = 2\kappa$. This can be used to combine Eqs. (\ref{eq:rot2})
and (\ref{eq:Dustrot}) into an algebraic equation for $\kappa^2$:
\begin{equation} \label{eq:rigrotkappa}
  \kappa^2 = (3{\mu} + 5p)/20 -2\omega^2/5 \ ,
\end{equation}
which, when differentiated (assuming an equation of state), yields
(i) $\kappa=0$, or (ii) $ 2{\mu} - 12\omega^2 - 3({\mu} + 
p){\mu}_{,p} = 0$. Case (i) will be treated for general vorticity in 
Sec. \ref{sec:ZeroKappa}.
For case (ii), the (inverse) generic solution becomes
\begin{equation}
  p(\mu) = 12\omega^2 - 3\mu + A|6\omega^2-\mu|^{3/2} \ ,
\end{equation}
where $A$ is a constant of integration. There is also the solution
$\mu=6\omega^2$. In general, we may choose $x^1=p$, which gives 
$y=2\kappa(\mu+p)$.

\subsection{Space-times with constant pressure}\label{sec:constpress}

From (\ref{eq:ContrBian}) we see that $a=0$. From Eq. (\ref{eq:Dustrot}) 
we then get $\omega^{2} = ({\mu} + 3p)/4$.
Differentiating this expression for $\omega$ and using the IC, we get
\begin{equation} \label{eq:rhodiff}
  y{\mu}' = 4\kappa({\mu} + 3p) \ .
\end{equation}
 
If the energy density is constant, we get a homogeneous model (see
Sec. \ref{sec:ST-Homogen}).
 
If the energy density is non-constant, we can choose
this as our spatial coordinate.
From (\ref{eq:rhodiff}) we see that $y = 4\kappa({\mu} + 3p)$. We then obtain
\begin{eqnarray*}
\fl  \kappa^2({\mu}) = ({\mu} + 3p)/2 + 2p + A({\mu} + 3p)^{1/2} \ , \\
\fl E({\mu}) = ({\mu} + 3p)/6 \ , \ \
    H^2({\mu}) = ({\mu} + 3p)^{2}/2 + A({\mu} + 3p)^{3/2} 
                  + 2p({\mu} + 3p) \ .
\end{eqnarray*}
 
Comparing with van Elst and Ellis \cite{vanElst} for dust space-times, we
see that it possible to integrate the system with the above choice of
coordinate, thus eliminating the procedure of solving the second order
equation (67) in \cite{vanElst} for $\mu$.

\subsection{Vanishing electric part of the Weyl tensor}

$E = 0$ is equivalent to the constraint $\omega^{2} - ({\mu} + 3p)/6 
+ a\kappa = 0$. Differentiating this (and using the IC) 
gives a constraint on the equation of state ${\mu} = {\mu}(p)$:
\begin{equation} \label{eq:constr}
  a({\mu} + p){\mu}_{,p} = 3(\kappa - a)({\mu} + 3p - 6a\kappa)
  \ .
\end{equation}
This equation can be used to replace (\ref{eq:rot1}), so that $\omega$
is determined by $E=0$.
From this we see that a constant energy density gives $a
=\kappa$, thus making the space-time conformally flat. But this implies
$\omega = 0$, contrary to our assumptions. 

The general system one obtains is fairly complicated, since it contains
$9^{th}$ degree polynomials in $a$ and $\kappa$, with coefficients consisting
of $p$, $\mu$ and its derivatives up to third order with respect to $p$. It
therefore seems unlikely that it is possible to find the general equation of
state.
A possible solution to this problem is to introduce
some further constraint, that will simplify the calculations.
We might insert a linear barotropic equation of state $p=(\gamma-1)\mu$.
The procedure above then gives a 
polynomial in $\gamma-1$, in which all coefficients are 
positive, which has to vanish. Thus we can conclude that
the only possible solutions have $\gamma<1$, i.e. these are all unphysical.

\subsection{Vanishing magnetic part of the Weyl tensor}

Now we have to impose the condition $a = \kappa$, so that
$\omega^{2} + a^{2} = ({\mu} + 2p)/3$. If we differentiate the
last expression, we obtain ${\mu}(p) = p + \mu_0$, where $\mu_0$ is some
integration constant. With $p$ as our coordinate, the
IC can then be integrated to give
\[
\fl  E(p) = (2p + \mu_0)/3 \  , \ \  \omega^2(p) = \omega_0^2 (2p
      + \mu_0) \  , \ \ a^2(p) = -\mu_0/6 + D(2p + \mu_0) \  . 
\]
This model belongs to a class described by Cahen and DeFrise in Ref.
\cite{Cahen}.

\subsection{Vanishing $\kappa$}\label{sec:ZeroKappa}

From (\ref{eq:rot2}) we have $\omega^{2} = ({\mu} + p)/2 $,
which, as in the preceding cases, can be differentiated, and it will yield
 ${\mu}(p) = -3p + \mu_0 $, if $a \neq 0$. Here $\mu_0$ is an 
integration constant. $E$ and $H$ are
give by Eqs. (\ref{eq:ElRot}) and (\ref{eq:MagRot}).
 
Choosing $p$ as our coordinate, we can integrate
Eq.~(\ref{eq:Dustrot}) if we insert the equation of state and the
expression for $\omega^2$. The result is
\begin{equation}
  a^2(p) = p + C/(-2p+\mu_0) ,
\end{equation}
where $C$ is an integration constant. In Ref. \cite{vanElst} there remained
a second order differential equation [Eq. (65)] to solve. This is eliminated
by the above choice of coordinate.

\section{Spatial twist in $\lowercase{e}_{\lowercase{1}}$: $\lambda \neq 0$}
 
Proceeding as in Case 1, we obtain the IC as the
differential equations
\begin{eqnarray}
  X\dot{{\be}}  & = & \lambda^{2} - ({\mu} + p)/2 +{\be}({\al} -
  {\be}) \, \, , \label{eq:lambda1} \\
  X\dot{{\al}}   & = & -2\lambda^{2} + ({\mu} - p)/2 -
  {\al}(2{\be} + {\al}) \, \, , \label{eq:lambda2} \\
  X\dot{\lambda} & = & \lambda({\al} - 2{\be}) \, \, ,
  \label{eq:lambda3} \\
  X\dot{{\mu}}    & = & -(2{\be} + {\al})({\mu} + p) \, \, , 
  \label{eq:lambda4}
\end{eqnarray}
where $\dot{}$ denotes differentiation with respect to $x^0$, 
and the algebraic relations
\begin{eqnarray}
  E & = & -2\lambda^{2} + 2{\mu}/3 - 2{\be}{\al}
  \label{eq:elpart} \, \, , \\
  H & = & 2\lambda({\al} - {\be}) \label{eq:Weyl} \, \, .
\end{eqnarray}
As in Sec. \ref{sec:Rotation}, we can reduce this system to three
differential equations for three functions in the generic case.

\subsection{Constant $\lambda$}

When $\dot{\lambda} = 0$, Eq. (\ref{eq:lambda3}) implies ${\al} =
2{\be}$. Inserting this into Eqs. (\ref{eq:lambda1}) and
(\ref{eq:lambda2}) gives
\begin{equation} \label{eq:lamal}
  {\be}^2 = -2\lambda^2/5 + (3{\mu} + p)/20 \ .
\end{equation}
This equation can be differentiated to give [by using the IC and
assuming $p = p({\mu})$] either (i) ${\be} = 0$ or 
(ii) $2({\mu}+p)p_{,{\mu}} - ({\mu}+3p) + 12\lambda^2 = 0$, where (i)
gives a homogeneous model.

As a simple ansatz for (ii) we might try $p = A{\mu} + B$, which gives 
(a) $p={\mu}+12\lambda^2$, or (b) $p=-{\mu}/2+3\lambda^2$.

\subsection{${\al} = c{\be}$, $c$ constant}

Within this class of models we have the shear-free and expansion-free
cases. 
Inserting the ansatz into the IC gives us 
\begin{eqnarray} 
  2(2c+1)c{\be}^2 + (2\lambda^2 - p - {\mu})c + 4\lambda^2 + p -
    {\mu} = 0 \ , \label{eq:alsquare} \\ 
  \left[{8(c + 1)}\lambda^2 + (p_{,{\mu}} - 1)({\mu} + p)c + 2({\mu} +
    p)p_{,{\mu}} -4p \right](c-1) = 0 \ . \label{eq:lamsquare}
\end{eqnarray}
We see that this equation is automatically satisfied if $c = 1$, i.e.,
if the fluid is shear free.
Finally, inserting Eq. (\ref{eq:lamsquare}) into Eq. (\ref{eq:lambda3}) 
gives us, with the help of the rest of the IC, the following second order 
differential equation for the equation of state:
\begin{eqnarray}
 && \left\{
    (c + 2)^2(\mu + p)\left[({\mu}+p){p_{,{\mu}{\mu}}} + 
    (p_{,{\mu}})^2\right] \right. \nonumber \\
 && \left.+ 2(c - 3)(c + 2)({\mu}+p){p_{,{\mu}}} - 3c^2({\mu}+p) +
    2c({\mu}-3p) + 16p \right\} \nonumber \\
  & \times& \left\{ (c + 2)^2({\mu}+p){p_{,{\mu}}} +
    6c({\mu}-p) + 4({\mu}-3p) \right\}(c-1) = 0 \label{eq:gigant}
\end{eqnarray}
Once again, the above equation is automatically satisfied if $c = 1$. 
Also, if $c=-2$ (i.e., the expansion free case), we get rid of all 
derivative terms in Eq. (\ref{eq:gigant}).
The remaining equation to solve is
(\ref{eq:lambda4}). One may make the ansatz $p=(\gamma-1){\mu}$ for the
equation of state. From Eq. (\ref{eq:gigant}) we then express $c$ in
terms of $\gamma$. There are three possibilities: (i) $\gamma(2+\gamma)c^2
+ 2(2\gamma^2-5\gamma+6)c + 4(2-\gamma)^2 = 0$, 
(ii) $c = 2(2-\gamma)/(2+\gamma)$, or (iii) $c=1$. 

Below we investigate some special values of $c$.

\subsubsection{$c=0$}
Since we have divided by $c$ at several places, we here use Eqs.
(\ref{eq:lambda1})-(\ref{eq:lambda4}) directly. The algebraic expression
for $\lambda$, which we obtain from Eq. (\ref{eq:lambda2}), is differentiated,
which gives $(\mu+p)p_{,\mu} + \mu - 3p = 0$, where an equation of state is
assumed. This has the solution
\begin{equation}
  2\mu/(\mu-p)+\ln|\mu-p|={\rm constant} \ .
\end{equation}
Choosing $\mu$ as the coordinate, we have
\begin{equation}
  \beta^2(\mu) = 
  e^{f(\mu)}\left(\frac14\int^{\mu}\frac{{\tilde\mu} + 3p({\tilde\mu})}
       {{\tilde\mu} + p({\tilde\mu})}e^{-f({\tilde\mu})}\d{\tilde\mu} 
       + A\right)
\end{equation}
where $A$ is a constant and $f(\mu) = \int^{\mu}\d{\tilde\mu}/[{\tilde\mu} 
+ p({\tilde\mu})]$.

\subsubsection{$c=1$, shear free case}
Equation (\ref{eq:gigant}) now becomes automatically zero. From
Eqs. (\ref{eq:elpart}) and (\ref{eq:Weyl}) we find that $E=H=0$.
Thus we have an $H_3$ and a $G_6$, which implies that we are dealing
with the open FLRW models. (Note that $\lambda$ is now redundant.)

Of the above the above equations, the only remaining are
Eqs. (\ref{eq:lambda3}) and (\ref{eq:lambda4}), while 
\begin{equation}
  {\be}^2 = -\lambda^2 +{\mu}/3 \ .
\end{equation}
Since we have no constraint on the equation of state, we choose ${\mu}$
as our coordinate (assuming it is non-constant). This implies, through
Eq. (\ref{eq:lambda4}), that 
$X = -3{\be}({\mu} + p)$. Equation (\ref{eq:lambda3}) can now be
integrated to yield
\begin{equation}
  \lambda({\mu}) = A\exp\left({\frac13\int^{{\mu}} 
     \frac{\d\tilde{{\mu}}}{\tilde{{\mu}} + p(\tilde{{\mu}})}}\right) \ ,
\end{equation}
(the case ${\be} = 0$ is space-time homogeneous). 
Here $A$ is some constant of integration.

As an example, we can solve this equation for $p = (\gamma -1){\mu}$,
which will give 
\[
\fl  \lambda = A{\mu}^{1/(3\gamma)} \ , \ \ \Theta^2 =
  -9A^2{\mu}^{2/(3\gamma)} + 3{\mu} \ , \ \ X^2 =
  3\gamma{\mu}\left[ -3A^2{\mu}^{2/(3\gamma)} + {\mu} \right] \ .
\]
Here we have used $\Theta = 3{\be}$.

\subsubsection{$c=-2$, expansion free case}
We have $\dot{\mu}=0$, and it is then straightforward to show that 
this is a homogeneous model.
The homogeneity of the model occurs because of our assumption of an
equation of state. We might take another route by not assuming this. We
then have 
\begin{eqnarray} 
  {\be}^2 = -({\mu} + 3p)/12 \ , \\
  X\dot{p} = 8{\be}\left[({\mu}-p)/4 - \lambda^2\right] \ .\label{eq:pequation}
\end{eqnarray}
Equation (\ref{eq:lambda3}) takes the form $X\dot{\lambda} =
-4{\be}\lambda$. We choose $\lambda$ as our coordinate. Then we can
integrate Eq. (\ref{eq:pequation}):
\begin{equation}
  p = {\mu} - 2\lambda^3/5 - C\sqrt{\lambda} \ ,
\end{equation}
where $C$ is some constant. All other quantities are then given in
terms of $\lambda$:
\[
\fl  E   = -2\lambda^2 + 2{\mu}/3 - \left(4{\mu} - 6\lambda^3/5 - 
        3C\sqrt{\lambda}\right)/3 \ , \ \ 
  H^2 = 3\left(4{\mu}-6\lambda^3/5-3C\sqrt{\lambda}\right)\lambda^2 \ .
\]

\subsection{Vanishing electric part of the Weyl tensor}
 
We now have the constraint $E=0$ $\Leftrightarrow$ $\lambda^{2} - {\mu}/3 
+ {\be}{\al} = 0$. 
Differentiating $E=0$ and using Eq. (\ref{eq:lambda3}),
 we get two cases: (i) ${\be} = {\al}$, i.e., 
no shear, or (ii) $5{\mu} - p = 18{\be}{\al}$.      
Since case (i) has already been treated above, we concentrate on case
(ii). From now on we assume $\dot{\mu} \neq 0$.
Using (ii) and the IC leads to the equations
\begin{eqnarray} 
  {\al} = \frac{2\left(1 - p_{,{\mu}}\right)}{3 +
    p_{,{\mu}}}{\be} \ , \label{eq:beta} \\ 
  {\be}^2 = \frac1{12}\frac{({\mu} + 3p)(1 - 2Q)}{2(1 +
    Q)Q_{,{\mu}} + Q(1 - 2Q)} \ , \label{eq:alfa}
\end{eqnarray}
where $Q \deq (1-p_{,{\mu}})/(3+p_{,{\mu}})$. But by using (\ref{eq:beta})
in the constraint (ii) we get ${\be}^2 = (5{\mu} -p)/(36Q)$. Thus we
obtain a constraint equation for the equation of state:
\begin{equation} \label{eq:stateconstr}
  16(5{\mu} - p)p_{,\mu\mu} +
  \left(3 + p_{,{\mu}}\right)\left(p_{,{\mu}} -
    1\right)\left(1 + 3p_{,{\mu}}\right)({\mu} -
    5p) = 0 \ .
\end{equation}

We can solve by making an
ansatz on the equation of state. We choose $p=(\gamma-1){\mu}$. Inserting
this into Eq. (\ref{eq:stateconstr}) gives $\gamma = 6/5$, and our equation of
state becomes $p={\mu}/5$. From Eqs. (\ref{eq:alfa}) and
(\ref{eq:beta}) we get
\begin{eqnarray*}
  {\be}^2 = 8{\mu}/15 \ , \qquad {\al}^2 = 2{\mu}/15 \ ,
\end{eqnarray*}
respectively.
This solution was found by Collins and
Stewart \cite{CollinsStew}.

\subsection{Vanishing magnetic part of the Weyl tensor}
 
From (\ref{eq:Weyl}) we get ${\be} = {\al}$, which implies $H=0$. 
This leads to
$X\dot{\be} = X\dot{\al}$ $\Rightarrow$ $\lambda^{2} -{\mu}/3 +
{\be}^{2} = 0$. But from (\ref{eq:elpart}) we see that $E = 0$, which implies
that this is the shear free case. Thus these are the open FLRW models.

\section{Homogeneous space-times} \label{sec:ST-Homogen}
 
This class of models is determined by algebraic equations only. 
We find that
$a=2{\be}+{\al}=0$ from Eq. (\ref{eq:contrBian}). 
Inserting this into the algebraic 
equations we deduce that $H=0$. Also, we obtain a number of very
simple relations such as
$\omega{\be}=\omega\kappa=\omega\lambda={\be}\kappa={\be}\lambda=
\lambda\kappa=E\kappa=(3E+{\mu}+p){\be}=0$, plus three equations
determining the relation between the kinematic and geometric
quantities. 
From these equations we get the solutions
\begin{enumerate}
  \item[1.] $\kappa = \lambda = \sigma = H = 0$, and $\omega^2 = 3E/2 =
    {\mu} = p$, which is G{\"o}dels model~\cite{Godel}.
  \item[2.] $\kappa = \lambda = \omega = H = 0$, and $3{\be}^2 = 3E/2 =
    -{\mu} = -p$, so that we have the restriction of stiff matter with
    negative pressure and energy density.
  \item[3.] $\kappa = \omega = {\be} = H = E = 0$, ${\mu} + 3p = 0$, and
    $\lambda^2 = {\mu}/3$, which is Einstein's static model. Since this
    has a $G_7$, the rotation coefficient $\lambda$ is redundant in
    this description.
  \item[4.] $\lambda = \omega = {\be} = H = E = 0$, ${\mu} + 3p = 0$, and
    $\kappa^2 = p$. This is once more Einstein's static model. Here,
    the rotation coefficient $\kappa$ is redundant.
\end{enumerate}

\subsection{Dependence on one null coordinate}

We can choose $x^0$ as our null
coordinate, so that $x = y = 0$ (since everything is symmetric
w.r.t. $x^0$ and $x^1$, it 
does not matter which one of them we choose). That $x^0$ is null means
that
$
  X = \ve Y
$,
where $\ve = \pm 1$. Next we insert this into the IC. This gives
$a = -\ve {\al}$ and $\omega = \ve \lambda$.

By using these equations in the IC it is straightforward to show that
these space-times are homogeneous, $\omega = \lambda = {\be} =
{\al} = E = H = \dot{{\mu}} = \dot{p} = 0$ and ${\mu} = -3p$ together
with $\kappa^2 = p$, i.e., this is Einstein's static model.

\section{The isometry algebra}\label{sec:Isometry}
 
With the method described above it is not only possible to find the
dynamics of the fluid, but it is also possible to find the metric in
the specific cases of interest. One uses the structure of the isometry
group on the orbits in $F(M)$ to make a metric ansatz. This is an easy
task in the case of three dimensional groups, since one then uses the
Bianchi classification, and it also works well in the four dimensional
case. From this one
can calculate the reduced set $S$ and compare with the original set
and from this obtain the metric components in terms of the kinematic
quantities, fluid variables and rotation coefficients.
 
We start out with Cartan's equations (\ref{eq:Cartan1}) and
(\ref{eq:Cartan2}), and project
these onto the orbits of the isometry group in $F(M)$. Working in
$F(M)$ has the advantage of always giving an isometry group acting
simply transitive \cite{KarlMac}. This in turn means that the
differential algebra between the
basis 1-forms (Cartan's equations) generates the structure constants
of the isometry group (if we choose an invariant basis $\{{\bfe}_I\}$),
i.e., if $\{{\bfxi}_P\}$ are the Killing vectors in $F(M)$ then 
\begin{eqnarray}
  \left[ {\bfxi}_P, {\bfxi}_Q \right] =  
         {\tilde C}^R\!_{PQ}{\bfxi}_R \  \Leftrightarrow  \
  \d{\w}^P| = \frac12 {\tilde C}^P\!_{RS} {\w}^R| \wedge {\w}^S| \ ,
  \label{eq:CartanProj} 
\end{eqnarray}
where ${\w}^P|$ is the projection onto the cotangent space of the orbits in 
$F(M)$ and the ${\tilde C}^P\!_{RS}$ are defined in 
terms of $C^I\!_{JK}$ through \cite{KarlBrad}
\begin{equation}
  {\tilde C}^P\!_{RS} = C^P\!_{RS} + C^P\!_{AB}
  I^A\!_{\bf c} I^{\bf c}\!_{|R} I^B\!_{\bf b} I^{\bf b}\!_{|S}
  - 2C^P\!_{AS} I^{A}\!_{\bf c} I^{\bf c}\!_{|R}\ .
\end{equation}
Thus the rotation coefficients and
the Riemann tensor [on the orbits in $F(M)$] essentially corresponds
to the structure constants of the isometry group of the space-time.
 
The orbits in $F(M)$ are defined as $\d I^{\bf a} = 0$, or when
using a fixed frame $\d x^{\alpha} = 0$, 
${\mbox{\boldmath $\tau$}}^a\!_b =
0$. This means that some of the 1-forms will be linearly dependent on 
the others. In the case of LRS, and with our
choice of frame, this means that, in the generic case, ${{\w}^{2}}_{3}$
is the connection 1-form that it linearly independent of the cotangent
basis.

In the following sections we find the metrics for the first two LRS 
classes, and discuss their relations to the IC.

\section{The isometry algebra for cases 1 and 2}\label{sec:Isom12}
  
The isometry algebra [Eq. (\ref{eq:CartanProj})] (when $\omega\neq 0$ or 
$\lambda\neq 0$) can be written as
  \begin{eqnarray}
    \d{\bfs}^1 &=& \Gamma{\bfs}^2\wedge{\bfs}^3 \ , \label{eq:genalg1}\\
    \d{\bfs}^2 &=& {\bfs}^3\wedge{\bfs}^4 \ , \label{eq:genalg2}\\
    \d{\bfs}^3 &=& {\bfs}^4\wedge{\bfs}^2 \ , \label{eq:genalg3}\\
    \d{\bfs}^4 &=& \Sigma{\bfs}^2\wedge{\bfs}^3 \ , \label{eq:genalg4} 
  \end{eqnarray}
where ${\bfs}^{\bf A} \deq {\w}^{\bf A}|$, ${\bf A}=2,3$,
and the definitions of the remaining quantities can be found in Table 1.

\begin{table}\label{1}
\caption{The definitions of the quantities used in Secs. 
       \protect\ref{sec:Isom12} and \protect\ref{sec:41}. 
       Notice that the main difference between the two
       cases lies in the causal properties of the essential coordinate $q$.}
   {\footnotesize{\rm
    \begin{tabular}{@{}lllllllll}
      \br 
      Case & ${\bfs}^1$  & ${\bfs}^4$                & ${\bfs}$
      & ${\bfs}_c$ & $\Gamma$   & $\Sigma$ & $q$ & $\upsilon$ \\ 
      \mr  
      1. $\omega\neq0$    & ${\w}^0|$ & $\omega{\w}^0|+{\w}^2\!_3|$ & 
      ${\w}^0$ & ${\w}^1$ 
      & $-2\omega$ & $E+{\mu}/3-3\omega^2+\kappa^2$ & $x^1$ & time-like \\
      \ms
      2. $\lambda\neq0$    & ${\w}^1|$ & $\lambda{\w}^1|+{\w}^2\!_3|$ & 
      ${\w}^1$ & ${\w}^0$ 
      & $2\lambda$ & $E+{\mu}/3+3\lambda^2-{\be}^2$ & $x^0$ & space-like
      \\
      \br
    \end{tabular}}}
\end{table}
  
We may regard Eq. (\ref{eq:genalg1}) as a
separate (differential) equation for ${\bfs}^1$, and Eqs.
(\ref{eq:genalg2})-(\ref{eq:genalg4}) as the algebra we need
to solve. ${\w}^2\!_3|$ can then be found through ${\bfs}^1$.
When $\omega\neq 0$, we know that ${\w}^1 = y^{-1}\d x^1$, and when
$\lambda\neq 0$ we know that ${\w}^0 = X^{-1}\d x^0$.

The isometry algebra can be split into three cases, according to
whether $\Sigma > 0$, $=0$ or $<0$.

\subsection{$\Sigma \neq 0$}

If we `normalize' our 1-forms as $\tilde{{\bfs}}^{{\bf A}} =
\sqrt{\Sigma}{\bfs}^{{\bf A}}$, 
Eqs. (\ref{eq:genalg2})-(\ref{eq:genalg4}) obtain the
structure of Bianchi type IX. The canonical solution is (see
Ref. \cite{RyanShep}) 
\begin{eqnarray}
  \tilde{{\bfs}}^{2} & = & -\sin\zeta \d\theta + \sin\theta\cos\zeta \d\phi
  \ \ , \label{eq:sphsym1} \\
  \tilde{{\bfs}}^{3} & = & \cos\zeta \d\theta + \sin\theta\sin\zeta \d\phi
  \ \ , \label{eq:sphsym2} \\
  {\bfs}^{4} & = & \cos\theta \d\phi + \d\zeta \ \ \label{eq:sphsym3}
  \ ,
\end{eqnarray}
where $\theta,\ \phi$, and $\zeta$ are some 
coordinates. The equation for ${\bfs}^1$ becomes $\d{\bfs}^1 =
(\Gamma/\Sigma)\tilde{{\bfs}}^2\wedge\tilde{{\bfs}}^3$. With the result above, we
can solve for ${\bfs}^1$ and obtain  
\begin{equation}
  {\bfs}^1 = (\Gamma/\Sigma)\cos\theta\d\phi \ ,
\end{equation}
where we have neglected a total differential in the integration
procedure. We may proceed in two ways according to the properties of
$\Sigma$. 

If $\Sigma > 0$, the coordinates in the solution are
real valued, and we can apply it directly. 

If $\Sigma < 0$, the new 1-forms introduced above take
their values 
over the set of complex functions. Thus we need to make the
transformation $\theta \rightarrow i\theta$, where the `new' $\theta$
is real. This will introduce an overall $i$ in the solutions for the
1-forms (since $\sin\theta \rightarrow i\sinh\theta$), which cancels
the $i$ in the definition of the `normalized' 1-forms. Also, in the
solution for ${\bfs}^1$, $\cos\theta \rightarrow \cosh\theta$, due to the
transformation.

\subsection{$\Sigma = 0$}

We can now define ${\bfs}^{\pm} \deq {\bfs}^2 \pm i{\bfs}^3$ and ${\bfs}^{\times} \deq
-i{\bfs}^4$. Equations (\ref{eq:genalg2})-(\ref{eq:genalg4}) then becomes
  \begin{eqnarray}
    \d{\bfs}^+        &=& {\bfs}^+\wedge{\bfs}^{\times} \ , \\
    \d{\bfs}^-        &=& -{\bfs}^-\wedge{\bfs}^{\times} \ , \\
    \d{\bfs}^{\times} &=& 0 \ .
  \end{eqnarray}
This has the structure of Bianchi type VI, and has the solution
\begin{eqnarray}
  {\bfs}^{+}      = e^{-\xi}\d\eta \ , \qquad
  {\bfs}^{-}      = e^{\xi}\d\tau \ , \qquad
  {\bfs}^{\times} = \d\xi \ .
\end{eqnarray}
Since the original 1-forms takes their values over the real functions,
we make the transformation $\xi = -i\zeta$, $\eta = \tau^{*} = \theta
+ i\phi$, for some real coordinates $\zeta$, $\theta$, and
$\phi$ (here $^*$ denotes complex conjugation). Thus we can write
  \begin{eqnarray}
    {\bfs}^{2} &=& \cos\zeta\d\theta - \sin\zeta\d\phi \,\,\, , \\
    {\bfs}^{3} &=& \sin\zeta\d\theta + \cos\zeta\d\phi \,\,\, , \\
    {\bfs}^{4} &=& \d\zeta  \,\,\, .
  \end{eqnarray}
The equation for ${\bfs}^1$ becomes $\d{\bfs}^1 =
(i/2)\Gamma{\bfs}^+\wedge{\bfs}^-$, which has the solution
\begin{equation}
  {\bfs}^1 = \Gamma\theta\d\phi \ .
\end{equation}

\section{Metrics for cases 1 and 2}\label{sec:41}
 
We here make the ans{\"a}tze for the full metric for space-times where
$\omega\neq 0$ or $\lambda\neq 0$.
When $\Sigma \neq 0$ the simplest guess is to take
\begin{eqnarray}
  {{\bfs}}            = \Sigma\chi(q){\bfs}^1 + \psi(q)\d\upsilon \ , \\ 
  {\w}^{{\bf A}} = \delta(q){\tilde{\bfs}}^{{\bf A}} \ .
\end{eqnarray}
When $\Sigma = 0$, we can make an ansatz analogous to the one above:
\begin{eqnarray}
  {\bfs}              = \chi(q){\bfs}^1 + \psi(q)\d\upsilon \ , \\
  {\w}^{{\bf A}} = \delta(q){\bfs}^{{\bf A}} \ .
\end{eqnarray}
(See Table 1 for the definitions of quantities introduced
above.)

\subsection{Metrics}

Introducing the parameter $\epsilon$, which is defined as $-1$ for
case 1, and $1$ for case 2, we can write the metric ${\met} \equiv 
\epsilon({\bfs}_c)^2 - ^{(3)}\!\!{\met}$, where ${\bfs}_c$ is the `complement'
to ${\bfs}$ (see Table 1), in one common form for all the
cases. The metric on surfaces $\{\d q = 0\}$ becomes
\begin{eqnarray}
  {}^{(3)}\!{\met} &= \epsilon({\bfs})^2 + ({\w}^2)^2 + 
      ({\w}^3)^2 \nonumber \\ 
     &= \left\{ 
    \begin{array}{ll}
      \vspace{2mm}
      \epsilon\left(\psi\d\upsilon +
        \chi\Gamma\cos\theta\d\phi\right)^{2} +
      \delta^{2}\left(\d\theta^{2} +
        \sin^{2}\theta\d\phi^{2}\right) \ ,& \Sigma>0 \\
      \vspace{2mm}
      \epsilon\left(\psi\d\upsilon +
        \chi\Gamma\theta\d\phi\right)^{2} +
      \delta^{2}\left(\d\theta^{2} + \d\phi^{2}\right) \ ,& \Sigma=0 \\
      \epsilon\left(\psi\d\upsilon +
        \chi\Gamma\cosh\theta\d\phi\right)^{2} +
      \delta^{2}\left(\d\theta^{2} +
        \sinh^{2}\theta\d\phi^{2}\right) \ ,& \Sigma<0 
    \end{array} 
  \right. \nonumber \\
  &= \epsilon\left[\psi\d\upsilon +
  \chi\Gamma\theta^{(1-s^2)}\left(e^{\sqrt{s}\theta} +
      e^{-\sqrt{s}\theta}\right)\d\phi\right]^2  \nonumber \\
  & \ \ + \delta^2\left[\d\theta^2 + 4^{-s^2}\left(e^{\sqrt{s}\theta} -
      s^2e^{-\sqrt{s}\theta}\right)^2\d\phi^2\right] . \label{eq:genmet}
\end{eqnarray}
Here we have introduced another parameter, $s$, which is defined
as $-1$ when $\Sigma>0$, $0$ when $\Sigma=0$, and $1$ when $\Sigma<0$.

\subsection{Consistency equations}
 
With consistency equations we mean a comparison between
the old set $S$ and the new set $\tilde S$ calculated from the metric
ansatz. As an explicit example of how to do this, we will look at case
1.

Starting from our metric ansatz, we use the type of tetrad that lead to the
ansatz (in this case a Lorentz-tetrad) and calculate
the rotation coefficients ${\tilde \Gamma}^a\!_{bi}$. We then observe that
  ${\tilde\Gamma}_{023} = \omega\chi/\delta^{2}$, which then gives
  $\chi = -\delta^{2}$ by comparison.
Inserting this into ${\tilde\Gamma}_{013}$ and comparing with
$\Gamma_{013}$ from $S$, we get the relation 
$\omega\delta^{2}/\psi={\rm constant}$. This constant can be
absorbed by redefining our time coordinate by a simple scaling (and
reflection). Doing this we obtain $\psi = \omega\delta^{2}$,
i.e., the only function from the ansatz that remains to be determined 
is $\delta$ (the others are known from the IC or their relation to $\delta$).

For case 1, the remaining equations are 
 \begin{eqnarray}
\fl  a = -y(\omega\delta^2)'/(\omega\delta^2) , \\  
\fl  \kappa = -y\delta'/\delta , \label{eq:del1} \\
\fl    E - ({\mu}+3p)/6 = -y\left[ (y\omega')'/\omega +
      2(y\delta')'/\delta \right] 
     - 4y^2\delta'\omega'/(\delta\omega) - 2(y\delta'/\delta)^2 , \\
\fl    -E/2 - ({\mu}+3p)/6 = -\omega^2 -
    y^2\delta'\omega'/(\delta\omega) - 2(y\delta'/\delta)^2 , \\
\fl    E/2 - {\mu}/3 = y(y\delta')'/\delta , \\
\fl    -E - {\mu}/3 = -3\omega^2 + s/\delta^2 +
    (y\delta'/\delta)^2 , \label{eq:del2} \\
\fl    H = -2y(\delta\omega)'/\delta \ ,
  \end{eqnarray}
where $s$ was introduced in the general metric (\ref{eq:genmet}).
 
If one insert the expressions for $a$ and $\kappa$ into the remaining
consistency equations, and then use the IC, one discovers that
the only equation that is not trivially satisfied when $s\neq0$ 
is Eq. (\ref{eq:del2}). Thus this equation
determines the metric when $s\neq 0$, ones the IC are solved. If
$s=0$, Eq. (\ref{eq:del1}) determines $\delta$. Case 2 can be treated in 
the same manner, and we give the function $\delta$ for the different 
cases in Table 2.

\begin{table}\label{2}
\caption{The form of the function $\delta$ for the 
   different cases. Here $A$ is an integration constant.}
  \begin{indented}
    \item[]\begin{tabular}{@{}lll}
      \br 
      Case & $\Sigma \neq 0$ & $\Sigma = 0$ \\
      \mr  
      1. $\omega\neq0$    & $\delta^2 = s(-E-\mu/3+3\omega^2-\kappa^2)^{-1} 
                = |\Sigma|^{-1} > 0$ 
           & $\delta = A\exp\left[-\int(\kappa/y)\d x^1\right]$ \\
      \ms
      2. $\lambda\neq0$    & $\delta^2 = s(-E-\mu/3+3\lambda^2+{\be}^2)^{-1} 
                = |\Sigma|^{-1} > 0$
           & $\delta = A\exp\left[\int({\be}/X)\d x^0\right]$ \\
      \br
    \end{tabular} 
  \end{indented}
\end{table}

The metric (\ref{eq:genmet}) can now be written as
\begin{eqnarray}
  {}^{(3)}\!{\met} &=
  \epsilon\left(\Gamma\delta^2/2\right)^2\left[\d\upsilon + 
     2\theta^{(1-s^2)}\left(e^{\sqrt{s}\theta} +
      e^{-\sqrt{s}\theta}\right)\d\phi\right]^2 \nonumber \\
  & \ \ + \delta^2\left[\d\theta^2 + 4^{-s^2}\left(e^{\sqrt{s}\theta} -
      s^2e^{-\sqrt{s}\theta}\right)^2\d\phi^2\right] \ ,
\end{eqnarray}
where the function $\delta$ is given in Table 2. Having solved the IC, we
know the metric explicitly, since all functions are determined, including
$\delta$.

\section{The function $\Sigma$ and its relation to the IC}\label{sec:Funs}

Since we have three different metrics depending on the properties of the 
function $\Sigma$, we must, through the IC, get some restriction on the 
constants of integration that has been introduced when solving the IC.
When $\mathrm{sgn}(\Sigma) = \pm 1$ we will get some inequality that 
have to be fulfilled
by the integration constants. This has to be treated separately for every 
solution of the IC. On the other hand, when $\Sigma = 0$, we have an 
algebraic constraint on the kinematic quantities. This will lead to more severe
restrictions than the inequalities impose. We can therefore treat this case on
its own.

\subsection{Case 1, $\omega \neq 0$}
Now we can write $0=\Sigma=E+\mu/3-3\omega^2+\kappa^2 = 
-p+2a\kappa-\omega^2+\kappa^2$. 
Taking the derivative of this and using the
IC, gives us either (i) $a=0$, or (ii) $\kappa=0$.

Case (i) was treated in Sec. \ref{sec:constpress}, but now there is the extra
equation $\omega^2 = -p+\kappa^2$. The energy density must not be constant, 
since then $\omega = 0$, i.e. a contradiction to our starting assumption. 
If the energy density is non-constant, we choose it as coordinate.
We then obtain the solution of Sec. \ref{sec:constpress}, but because of our
extra constraint $\Sigma =0$, we find that the integration constant $A=0$.

For (ii), we obtain $\omega^2=-p$, so that the pressure is negative. We find
that $p$ must not be constant, because then $\omega=0$.
Thus this case is equivalent to the one treated in Sec. \ref{sec:ZeroKappa}
with the constant of integration $\mu_0=0$. 

\subsection{Case 2, $\lambda\neq0$}
The constraint becomes $0 = \Sigma = E + \mu/3 + 3\lambda^2 - {\be}^2
= \lambda^2 + \mu - {\be}({\be} + 2{\al})$. Differentiating this and 
using the IC gives an identity. Thus we can get rid of, say, $\lambda$ 
(expressing it algebraically in terms of the other functions) and get a 
simpler set of integrability conditions.
The IC may then be solved (as before) for some special cases, 
e.g. an expansion free fluid.

\section{Concluding remarks}
 
As a conclusion one might point on the advantages of this method
compared with other methods. First, we automatically obtain a reduced
set of equations (although they may contain redundant information)
needed to be solved to find a solution to Einstein's equations. Second,
we have a split in the equations, one set consisting of the equations
mentioned above, and one set that determines the rest of the metric
[the `redundant' part connected to the symmetry of $(M,{\met})$].
We hope that these examples have shown on some of the strengths of the
method presented.

\ack

The author would like to thank Michael Bradley for support and
enlightening discussions, and Claes Uggla and Ulf Nilsson at 
Stockholm University for helpful comments. 
Also, during the writing of this paper, REDUCE 
and Jan {\AA}mans' CLASSI \cite{AAman} were valuable tools.

\appendix
 
\section{Integrability conditions}

In our fixed frame formalism we get the following IC:
\begin{enumerate}
\item[1.] Commutator equations 
  ${x^{\alpha}}_{|[k,|{\beta}|}{x^{{\beta}}}_{|l]} = {x^{\alpha}}_{|m} 
{{\Gamma}^{m}}_{[kl]}$, with the indices
$\left(^{\alpha}_{kl}\right)$ given in the bracket:
 \begin{eqnarray}
   \fl \left(^0_{01}\right) \qquad &  
  \dot{X}Y + X'y - \dot{Y}X - Y'x  
   =  Xa + Y{\al} \ , \label{eq:commcoord1}\\ 
   \fl \left(^0_{23}\right) & 
   0 = X{\omega} - Y{\lambda} \ , \label{eq:class1} \\ 
   \fl \left(^1_{01}\right) & 
   \dot{x}Y + x'y -  \dot{y}X - y'x   
   = xa + y{\al},\label{eq:commcoord2}\\ 
   \fl \left(^1_{23}\right) & 
   0 = x{\omega} -  y{\lambda}, 
   \label{eq:class2} 
  \end{eqnarray}
\item[2.] Ricci identities 
  ${R^{a}}_{bij} = 2{{\Gamma}^{a}}_{b[j,|\alpha |}{x^{\alpha}}_{|i]} + 
  2{{\Gamma}^{am}}_{[j}{{\Gamma}}_{|bm|i]} +
  2{{\Gamma}^{a}}_{bk}{{\Gamma}^{k}}_{[ij]}$, with the indices
  $(^a_{bij})$ given in the bracket:
  \begin{eqnarray}
   \fl \left(^0_{101}\right) \qquad & Y\dot{a} + ya' 
   =  E +a^{2} -({\mu} + 3p)/6 
   - {{\al}}^{2} - X\dot{{\al}} - x{{\al}}'  , \label{eq:ay}\\
   \fl \left(^0_{123}\right) & H =  2(a - {\kappa}){\omega} 
   + 2({{\al}} - {{\be}})\lambda ,  \label{eq:H} \\
   \fl \left(^0_{202}\right) & X\dot{{\be}} +
   x{{\be}}'  = - 
   ({\mu} + 3p)/6 - {{\be}}^{2} - E/2 + a{\kappa}
   + {\omega}^{2}, \label{eq:alfax}\\
   \fl \left(^0_{212}\right) & Y\dot{{\be}} +
   y{{\be}}' =  
   {\kappa}({{\be}} - {{\al}}) + {\lambda}{\omega}  , \label{eq:alfay}\\
   \fl \left(^0_{203}\right) & X\dot{\omega} +
   x{\omega}' = -a{\lambda} - 2{\omega}{{\be}}  , \label{eq:B}   \\
   \fl \left(^0_{213}\right) & Y\dot{\omega} +
   y{\omega}' = -H/2 + {\kappa}{\omega} + 
   {\lambda}({{\al}} - {{\be}})  , \label{eq:A}  \\
   \fl \left(^1_{202}\right) & X\dot{\kappa} +
   x{\kappa}' =  (a - {\kappa}){{\be}} - 
   {\lambda}{\omega}  , \label{eq:E}  \\
   \fl \left(^1_{212}\right) & Y\dot{\kappa} +
   y{\kappa}' = -E/2 + {\kappa}^{2} - 
   {\lambda}^{2} + {\mu}/3  - {{\be}}{{\al}}, \label{eq:kappa2}\\
   \fl \left(^1_{203}\right) & X\dot{\lambda} +
   x{\lambda}' =  -(a - {\kappa}){\omega} + 
   H/2 - {\lambda}{{\be}}, \label{eq:lambdaA} \\
   \fl \left(^1_{213}\right) & Y\dot{\lambda} +
   y{\lambda}' = 2{\kappa}{\lambda} + 
   {\omega}{{\al}} , \label{eq:lambdaB} 
  \end{eqnarray}
\item[3.] Bianchi identities 
  ${R^{p}}_{q[ij;k]} =  0 $, with $\{^p\!_q\} = \{^2\!_3\}$, and the
  indices $(_{ijk})$ given in the bracket: 
 \begin{eqnarray}
   \fl (_{023}) \qquad & X{\left(\dot{E} + \dot{{\mu}}/3 \right)} 
   + x{\left( E' + \mu'/3 \right)} & =  
   -(3E + {\mu} + p){{\be}} - 3H{\lambda} , \label{eq:C}  \\
   \fl (_{123})        & Y{\left( \dot{E} + \dot{{\mu}}/3 \right)} 
   + y{\left( E' + \mu'/3
     \right)} & =  3E{\kappa} - 3H{\omega} . \label{eq:D}
 \end{eqnarray}
\end{enumerate}
Here we have used the notation $\dot{} \, =
{\partial}/{\partial}x^{0}$ and 
$ ' \, = {\partial}/{\partial}x^{1}$.
 
We observe that the cyclic identities ${R^{t}}_{[ijk]}   =  0$ are
already imposed by our choice of Riemann tensor.

\end{document}